\documentclass[conference]{IEEEtran}
\IEEEoverridecommandlockouts
\usepackage{fancybox}
\usepackage[most]{tcolorbox}
\usepackage{url}
\usepackage[backend=bibtex,style=ieee,natbib=true,maxcitenames=2, mincitenames=1]{biblatex}

\usepackage{amsmath,amssymb,amsfonts}
\usepackage[ruled,linesnumbered, noend, noline]{algorithm2e}
\usepackage{varwidth}
\usepackage{graphicx}
\usepackage{textcomp}
\usepackage{enumitem}
\usepackage{mathtools}
\usepackage{xcolor}
\usepackage[hidelinks]{hyperref}
\usepackage[capitalise, nameinlink]{cleveref}
\usepackage{orcidlink}
\usepackage{listings}
\usepackage{relsize}
\usepackage{siunitx}
\usepackage{xspace}
\usepackage{balance}
\usepackage{blindtext}
\usepackage[T1]{fontenc}
\usepackage{tabularx}
\usepackage{booktabs}

\definecolor{eclipseBlue}{RGB}{42,0.0,255}
\definecolor{eclipseGreen}{RGB}{63,127,95}
\definecolor{eclipsePurple}{RGB}{127,0,85}
\definecolor{lightblue}{RGB}{58, 125, 201}
\definecolor{stringgreen}{RGB}{35, 107, 51}
\definecolor{verylightgray}{RGB}{248, 248, 248}

\def\BibTeX{{\rm B\kern-.05em{\sc i\kern-.025em b}\kern-.08em
    T\kern-.1667em\lower.7ex\hbox{E}\kern-.125emX}}
    
\crefname{lstlisting}{listing}{listings}
\Crefname{lstlisting}{Listing}{Listings}

\newlength{\mylength}
\makeatletter
\newcounter{ObservationBoxCounter}
\newenvironment{ObservationBox}{
\par
\refstepcounter{ObservationBoxCounter}
\protected@edef\@currentlabelname{Observation \theObservationBoxCounter}
\protected@edef\@currentlabel{Observation \theObservationBoxCounter}
\vspace{0.2cm}\noindent%
\setlength{\fboxsep}{0.3cm}%
\setlength{\mylength}{0.985\columnwidth}%
\addtolength{\mylength}{-2\fboxsep}%
\addtolength{\mylength}{-2\fboxrule}%
\begin{Sbox}
\begin{minipage}{\mylength}%
  }{
  \end{minipage}%
  \end{Sbox}%
  \cornersize{.1}
  \Ovalbox{\TheSbox}
  
}
\tcbset{
    frame code={},
    center title,
    left=0pt,
    right=0pt,
    top=0pt,
    bottom=0pt,
    colback=gray!25,
    colframe=white,
    width=\dimexpr\textwidth\relax,
    enlarge left by=0mm,
    boxsep=5pt,
    }

\usepackage{tikz}
\usetikzlibrary{shapes.geometric, calc}

\tikzset{modeoval/.style={ellipse, draw, font=\footnotesize,
  inner xsep=0pt, inner ysep=2.5pt, line width=0.5pt}}

\title{The Comprehension Gap in LLM-Assisted Software Development: An Experiment}

\author{Julian Oertel\,\orcidlink{0009-0000-5919-8638} \\
Software Engineering \\
University of Rostock \\
julian.oertel@uni-rostock.de
\and
Jil Klünder\,\orcidlink{0000-0001-7674-2930} \\
Software Engineering and Data Science \\
University of Applied Sciences \textbar\ FHDW Hannover \\
jil.kluender@fhdw.de
\and
Regina Hebig\,\orcidlink{0000-0002-1459-2081} \\
Software Engineering \\
University of Rostock \\
{regina.hebig@uni-rostock.de}
}
\date{}

\makeatletter
\newcounter{RQCounter}
\newenvironment{ResearchQuestion}{%
\par\nointerlineskip%
\refstepcounter{RQCounter}%
\protected@edef\@currentlabelname{RQ\theRQCounter}%
\protected@edef\@currentlabel{RQ\theRQCounter}%
\vspace{0.2cm}\noindent%
\begin{minipage}[b]{\columnwidth}%
\textbf{RQ\theRQCounter: }}{%
  \end{minipage}%
}%
\makeatother

\begin{document}

\maketitle
\begin{abstract}
When LLM tools are pushed beyond their capabilities for software development tasks, developers must be able to step in. However, the comprehension that developers acquire during development might be negatively impacted due to excessive code generation. This could create a cognitive debt that needs to be resolved before tasks can be solved efficiently again.
We explore how well developers understand code they generated by prompting in comparison to self-written and unseen code and how they reflect on their comprehension.
We conduct a crossover experiment and interviews with 35 participants consisting of students, academic personnel, and software developers. After completing two programming tasks with or without LLM chat usage, respectively, participants are tasked with finding introduced mutations in the produced and additional unseen code. We use insights from interviews and screen recordings to supplement our results.
Regarding the experiment, 36\% of participants report a better understanding of self-written code compared to their generated code. On average, participants solved the comprehension task 20\% faster for self-written code than for generated code and found 11\% more mutations.
Participants needed significantly less time to solve programming tasks when working with the LLM. However, participants with high reliance on the LLM chat and low task engagement showed significantly decreased comprehension.
Our qualitative results show great variability in individual perspectives on comprehension differences and strategies to mitigate issues.
Strategies for dealing with comprehension differences range from accepting comprehension differences to adapting their LLM interactions or avoiding the tools altogether. Our results warrant further investigation into cognitive debt in larger systems and developer practices to mitigate issues.

\end{abstract}

\section{Introduction}
There is an increasing use of Large Language Models (LLMs) in software projects\footnote{visible e.g., when comparing \url{https://survey.stackoverflow.co/2024/ai} and \url{https://survey.stackoverflow.co/2025/ai}}, for example, to support documentation, to produce source code, or to write test cases~\cite{sergeyuk}. LLMs achieve sufficient correctness for simple tasks, depending on the programming language, problem complexity, and prompt~\cite{OERTEL2025107737,liu2024no}.

However, when LLMs produce only partially correct results (if any), developers are required to step in. How efficiently and effectively they can do so depends on how well they understand that code, as the largest share of developers' time is spent on comprehension activities~\cite{xia2018}, and incomplete understanding makes modifications error-prone~\cite{LITTMAN1987341}.
Developers who write code themselves build their mental models along the way. For code produced with LLMs this does not need to be the case anymore. The time saved during generation may, therefore, only be deferred and has to be ``paid back'' once the code needs to be understood by a human in order to correct, extend, or verify it, a cost which can even surpass the initial speedup. However, this heavily depends on the abilities and adoption of LLM tools~\cite{becker2025measuringimpactearly2025ai}\footnote{cf. \url{https://metr.org/blog/2026-02-24-uplift-update/}}.

This has recently been described as cognitive debt~\cite{storey2026} or comprehension debt~\cite{ahmad2026}. In a review of grey literature, \textcite{tavantzis2026cognitivedebtgenaiera} find that cognitive debt can accumulate unnoticed, particularly when developers accept AI-generated solutions without building understanding. Practitioners report consequences such as harder maintenance and debugging, fragmented knowledge in teams, and reduced productivity~\cite{tavantzis2026cognitivedebtgenaiera}. 

The term ``Cognitive Debt'' remains loosely defined and is used for different related phenomena (see \Cref{sec:rel_work}). 
\textcite{tavantzis2026cognitivedebtgenaiera} adopt a working definition of cognitive debt as ``the accumulated gap between the software understanding available to an individual or team and the understanding required to perform current or future software engineering work''. In this paper, we focus on the individual level and define \emph{individual cognitive debt} as a person's insufficient degree of comprehension of a software system with regard to performing a specific task (cf.~\cite{wyrich2023sourcecodecomprehensioncontemporary}). 
People can experience consequences of cognitive debt, for example, when working on an unknown software system or on parts of a system that were not authored by them, or after forgetting particulars of a system. With LLMs, developers may also incur cognitive debt for code they have just produced themselves.


Empirical evidence on cognitive debt in software development is still scarce, and \textcite{tavantzis2026cognitivedebtgenaiera} call for studies that measure individual and team understanding. Existing experiments mostly assess developers' understanding of libraries or existing code bases after working with AI assistants~\cite{shen2026ai,qiao2026,Alakmeh2026}. We therefore investigate how well developers understand code they produced with LLM assistance compared to code they wrote themselves.
We present the results of an experiment with 35 participants in which participants are first asked to write and generate code according to a problem description, before they try to comprehend their code in addition to an unknown code snippet with the goal of finding mutations in the code. We extend the insights gained from the tasks with interviews in which we asked the participants to share their perspective on comprehension of generated code.


The remainder of the paper is structured as follows: \Cref{sec:rel_work} presents related work. In \Cref{sec:study_design}, we outline our study design and present our results in \Cref{sec:results}. In \Cref{sec:discussion}, we interpret our results and address possible threats to validity in \Cref{sec:threats}. We conclude the paper in \Cref{sec:conclusion}.



\section{Related Work}\label{sec:rel_work}
\subsection{Cognitive Offloading}
\textcite{Risko2016} define cognitive offloading as ``the use of
physical action to alter the information processing requirements of a task so as to reduce cognitive demand''. An example is the use of a calendar system to remind oneself of a particular event.

While offloading can result in positive and negative effects~\cite{Risko2016}, often there is a trade-off when choosing to offload cognition. For example, \textcite{Grinschgl2021} show the trade-off in three experiments with \(n\)=172 participants each: An increase in immediate task performance through cognitive offloading led to a reduced performance in a subsequent unexpected memory test. However, if participants were made aware of the memory test but forced to offload, negative effects almost vanished. \textcite{Dahmani2020} report on a study showing a negative impact on spatial memory after habitual use of GPS systems, highlighting the possible long-term effects of cognitive offloading.

With generative AI a new tool emerged for offloading and automating reasoning. The consequential ``ironies of automation'' have long been studied in human factors research and now affect GenAI users as they shift from producing artifacts to evaluating them~\cite{Bainbridge1983, Simkute04032025}. 

Recent studies show that delegating tasks to LLMs also affects understanding in software development and engineering.

\textcite{shen2026ai} find in a between-subjects experiment (\(n\)=52) that developers learning an unfamiliar Python library with AI assistance scored lower on a subsequent comprehension quiz, especially when they delegated code generation. \textcite{qiao2026} found that in a brownfield setting, GitHub Copilot increased task performance without improving comprehension of the code base (\(n\)=15), although frequent verification loops were strongly associated with higher comprehension. \textcite{Alakmeh2026} find no major comprehension differences between AI and no-AI conditions (\(n\)=21), but their persona taxonomy reveals that AI-steered developers performed well only with AI assistance and struggled with comprehension and transfer tasks.

\subsection{Cognitive and Comprehension Debt}
\textcite{Cunningham1992} coined the ``debt'' metaphor in 1992 for the trade-off between short-term gains in efficiency by shipping immature code for the risk of long-term quality issues bringing development to an (almost) standstill. 
With generative AI tools allowing the offloading of cognition, the concepts of ``Cognitive Debt'' or ``Comprehension Debt'' have gained traction. 

\textcite{Kosmyna2025} introduce ``Cognitive Debt'' in an EEG study (\(n\)=54) on essay writing. ChatGPT users showed the weakest neural connectivity, lowest perceived ownership, and remained less engaged when writing without the tool (observed for remaining \(n\)=18 participants in the last session).

In a survey of 299 STEM students, \textcite{CHOUDHURI2026108346} find that trust-driven, routine GenAI use is associated with less reflection, need for understanding, and critical thinking. They propose a cognitive debt cycle in which repeated delegation weakens students' ability to perform tasks independently, which in turn increases their reliance on GenAI. In these interpretations, cognitive debt describes a decline of individuals' mental engagement and skills.

\textcite{storey2026} uses ``Cognitive Debt'' as a team-level property that describes the erosion of shared understanding of a system. Together with intent debt and technical debt, it forms the triple-debt model to assess software health.

Participants of a focus group (\(n\)=22) by \textcite{vanheesch2026} view it as an organizational risk of accepting responsibility for parts of a system one does not fully understand, distinct from skill degradation.

Empirical studies on missing comprehension of concrete software artifacts remain rare. \textcite{ahmad2026} identifies accumulation patterns such as black-box code acceptance in 621 student diaries, and \textcite{chou2026vibeguidegraphbaseduserinterface} show that a graph-based interface can shift developers' comprehension of agent-generated projects towards relationships between components (\(n\)=16). In contrast, we measure developers' comprehension of code they produced themselves.




\subsection{Comprehension of Self-Written Code}
In the mapping study by \textcite{wyrich202340} of 95 code comprehension experiments, we found none that measured participants’ comprehension of their own code.

The comprehension of self-written code has been studied in computing education, mainly by asking questions about the code (Questions about Learners' Code)~\cite{Lehtinen2021Lets,Lehtinen2021,Lehtinen2023,Santos2022,Canico2026}. Outside education, memory and familiarity with one's own software systems over time has been studied by \textcite{kruger2018you} and \textcite{kruger2020developers} using self-assessed knowledge of developers~\cite{kruger2018you, kruger2020developers} and asking questions about the systems~\cite{kruger2020developers}.

\textcite{balepur2026} investigate the comprehension of students (\(n=54\)) of the code they produced using a coding agent or with a chat bot in a between-subjects study. They find that students who built a website with a coding agent scored substantially lower than chatbot users on recall and purpose questions about their own code.


\section{Study Design}\label{sec:study_design}
We conduct a crossover experiment to investigate differences in the comprehension developers gain of code when programming with LLM assistance compared to programming without such assistance. 

\begin{figure*}
    \centering
    \includegraphics[width=0.99\textwidth]{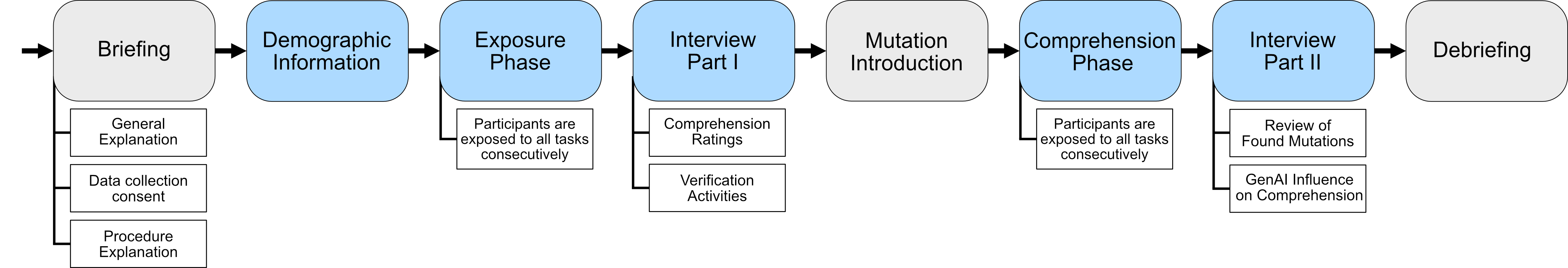}
    \caption{Overview of the study procedure. Blue steps indicate participant activities or active involvement.}
    \label{fig:procedure_overview}
\end{figure*}

A general overview of the experiment is given in \Cref{tab:expOverview}, and the experiment structure is given in \Cref{fig:procedure_overview}. In summary, the experiment consists of an exposure phase in which participants encounter three programming tasks in three different modes: Two modes, where participants solve a programming task, without LLM (\textsc{self}) or with LLM (\textsc{llm}), and one (\textsc{unseen}) mode, in which participants only get to read a description of a programming task, without actually solving it. Afterwards, in the comprehension phase, they are tasked with finding mutations in the previously produced code and the unseen code of the \textsc{unseen} mode. With semi-structured interviews in between phases, we capture participants' perspectives. 
\begin{table}[t]\centering
\caption{Experiment overview}\label{tab:expOverview}
\begin{tabularx}{0.9\columnwidth}{lX}\toprule
Goal & Investigating impact of using LLM tools for programming on cognitive debt\\
Independent variables &  Mode: \textsc{self}, \textsc{llm}, or \textsc{unseen} \\
Tasks & Write code, find mutations \\
Dependent variables & Found mutations, programming time, time in mutation finding task, self-assessed comprehension (description, solution) \\
Confounding factors & Professional programming experience, Programming proficiency, language proficiency, LLM proficiency for coding, order, task\\
Design & Crossover \\
\bottomrule
\end{tabularx}
\end{table}

\subsection{Research Questions}
We pose the following research questions:
\begin{ResearchQuestion}\label{rq1}
    How well do developers understand code they produced using generative AI compared to self-written code?
\end{ResearchQuestion}
\begin{ResearchQuestion}\label{rq2}
How do developers perceive their comprehension of code they produced using generative AI?
\end{ResearchQuestion}
\subsection{Participants}\label{subsubsec:participant_assignment}

\begin{figure}[t]
    \centering
    \includegraphics[width=0.9\linewidth]{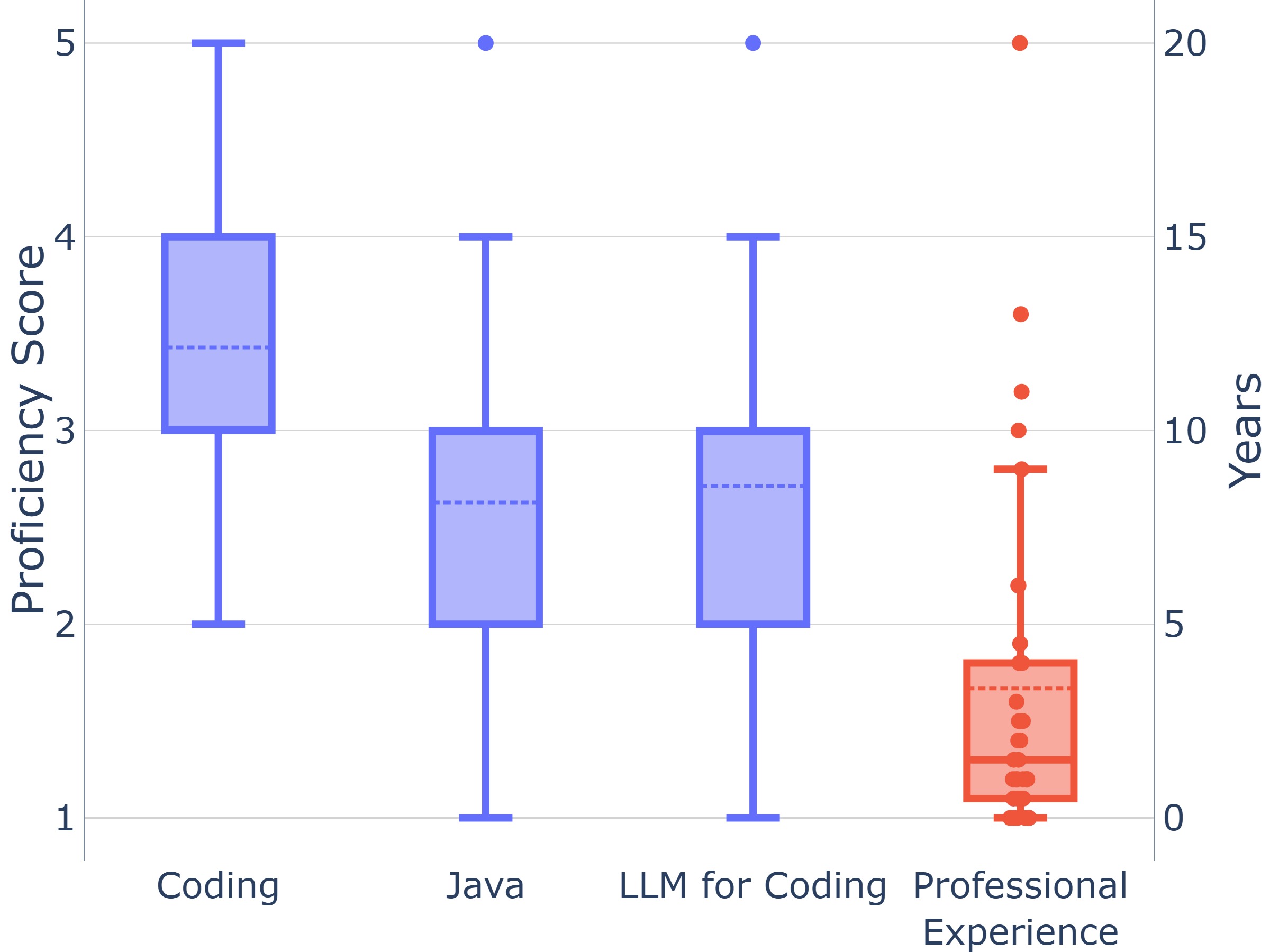}
    \caption{Box plot showing the self-reported proficiency of participants for general coding, coding with Java, and using LLMs for coding (this score was given based on frequency and experience of use) on a 5-point scale. Professional experience is measured in years.}
    \label{fig:participant_demographics}
\end{figure}

We applied sequential recruitment. That is, we did not know all participants beforehand but recruited them over the course of our study.
The inclusion criterion for participation was having basic programming skills with Java (e.g., demonstrated by having completed at least one university course in which Java was used) in order for participants to be able to complete the experiment in a reasonable amount of time. We used convenience sampling, drawing on industry contacts, and advertising to students in university courses. Our sample includes undergraduate and graduate students, PhD candidates, academic personnel, and professional software developers. 
We were able to compensate students with a 10\,€ gift voucher for their participation. We recruited 38 participants. However, due to deviations in the experiment (cf.~\Cref{{subsec:deviations}}) not every participant could be included in the analysis. We present an overview of participant information in \Cref{fig:participant_demographics}.

\subsection{Operationalization of Comprehension}
Following Wyrich~\cite{wyrich2023sourcecodecomprehensioncontemporary}, we understand comprehension as the ``degree of accomplishment in inferring the meaning of source code'', judged relatively to the task a developer pursues. We measure comprehension with the task of finding errors in code, which is one of the task types often used in code comprehension studies~\cite{wyrich202340}. Because errors are detected during the ongoing comprehension process~\cite{GILMORE1991151}, how many errors participants find, and how quickly, reflects the mental model they bring to the code.

What this mental model contains depends on the mode. In the \textsc{unseen} mode, participants have no prior model of the code. They must trace the code and judge it against the task description, which corresponds to the analytical and critical facets of \textcite{bartl_et_al}. In the \textsc{self} mode, participants designed the solution and iterated on it, so they can also compare the code against what they remember having written and why. The \textsc{llm} mode likely lies in between the two. Participants write prompts and judge and integrate the output, but do not construct the code themselves or only do so to a lower degree. We chose this task for three reasons: (1) we can keep the task identical across modes unlike pure recall or reproduction tasks; (2) judging is easy unlike free-text responses to questions; and (3) mutation operators can be applied easily and independently of the concrete solution.


We measure the time taken to find mutations and the number of correctly identified mutations for each task and participant. We complement this measurement with a comprehension rating of participants.
\subsection{Material}
\subsubsection{Programming Tasks}
We prepared three programming tasks that were complex enough to observe differences in comprehension, but also short enough to fit into our study procedure. In addition, we wanted the tasks to be somewhat diverse in the required competences and the required LLM interactions. Each of the tasks requires participants to implement one method only.

We built \textbf{Task 1} based on the method chomp() of the Apache Commons Lang library, which removes the last line break of a string. This task serves as a small algorithmic challenge and is usually very easily solvable using an LLM. \textbf{Task 2} is a ``Grade calculator'' that is supposed to output grades based on an achieved percentage. This task requires less algorithmic thinking. However, we observed that the LLM used (cf. \cref{subsubseq:llm_tool}) was consistently (10/10 attempts) not able to solve this task on the first attempt using only the task description, as it introduced type errors and used the wrong order of steps.
For \textbf{Task 3}, participants needed to implement an ``Invoice calculator'', applying different calculations based on customer type and order amount. This task involves multiple classes, which requires participants to familiarize themselves with the context of the method and provide this context to the LLM.

Each task consists of a title, a description of the desired behavior, and a usage example in a PDF file. We created IntelliJ projects for each task, providing a designated class containing the method signature and a main method. The main method is given to simplify the execution of the method for testing purposes. We also include a test suite consisting of two simple test cases that participants can freely alter and expand. The projects are reset before each new participant using Git. 

For the exposure phase, we created an additional more comprehensive test suite that is meant to be used by the experimenter to test the solutions created by the participants. 

For the purpose of the comprehension phase in the \textsc{unseen} mode, we prepared solutions for each task by extracting them from the Apache Commons Lang library (Task 1) or by generating a solution using an LLM (Tasks 2 and 3). To check their correctness, we utilized the same test suite as for the participants' solutions in the exposure phase. We then created mutated versions of these solutions by introducing five mutations using the mutation operators described in the following.

\subsubsection{Mutation Operators}
For the comprehension phase, mutations are manually introduced into the code by the experimenter. We opt to exclude operators that lead to syntax errors because we do not aim to focus on assessing language understanding. Instead, our focus is on assessing familiarity and understanding of the code logic and algorithmic structure. We therefore chose the following mutation operators inspired by the operators included in the MuJava library~\cite{ma2005mujava}: 
\begin{itemize}
    \item Off-By-One, i.e., adding or subtracting one to integer expressions
    \item Condition-Altering, e.g., negating boolean expressions or changing boolean operators
    \item Arithmetic-Altering, i.e., changing arithmetic operators (e.g., plus to minus) or reordering terms
    \item Enumeration-Value-Exchange, i.e., substituting enumeration values for others
    \item Control-Flow-Altering, e.g., turning else-if to if
\end{itemize}
When introducing the mutations, we ensure that the code can be compiled and causes semantic errors and runtime exceptions only. Also, we try to inject similar mutations for similar solutions of participants.

\subsubsection{LLM Tool}\label{subsubseq:llm_tool}
For the exposure task, we equip participants with a chat to interact with our locally hosted Q8\_0-quantized version of Qwen3-32B\footnote{\url{https://huggingface.co/Qwen/Qwen3-32B-GGUF}}. For hosting the model, we use Llama.cpp\footnote{\url{https://github.com/ggml-org/llama.cpp}}. We opted for a locally hosted model for data protection reasons.
We disable the thinking mode to provide faster feedback for participants and set the hyperparameters according to the LLM provider's recommendations (\textit{temp}=\(0.7\), \textit{TopP}=\(0.8\), \textit{TopK}=\(20\), \textit{MinP}=\(0\)). We validated the model and the configuration with the goal of ensuring that the provided tool is capable of helping participants to solve the tasks. Therefore, we used our tasks and interaction patterns that we expect from participants (such as asking for specific methods in Java, analyzing errors, or refactoring code). We could confirm that the tool was generally able to generate correct code for all three tasks if sufficient context was provided.

\subsection{Experiment Structure}

This section outlines the general procedure of the study, consisting of an experiment and a semi-structured interview.
The study is guided by a custom-built experiment application that bundles the individual steps and serves as a reference point for participants and the experimenter. The application is also used to collect participant information and assign conditions as well as to manually introduce mutations by the experimenter, and the participants use it to mark mutations.
The studies were conducted individually with each participant. We conducted the experiments locally or remotely, using Zoom or Teams. All tasks were solved on the computer of the experimenter directly or through remote access. 

\subsubsection{Briefing of Participants}
The initial briefing of the participants consists of a rough outline of the study procedure. We also make sure that participants can familiarize themselves with or alter the provided experiment setup (e.g., adjusting the mouse speed or changing the keyboard layout). For participants unfamiliar with IntelliJ, we gave a short introduction to executing code and test cases and to the usage of the debugger. 

We inform participants that one goal of our study is to investigate the impact of using generative AI on code quality and developer efficiency, and to investigate developer interactions and sentiment when working with LLM tools.
We require this indirection in order to delay the insight that the study is about code comprehension during the initial exposure phase to prevent participants from studying the code to an unusual degree (cf.\,\cite{Grinschgl2021}). 
For this reason, we do not yet mention the comprehension task to the participants and only give information on the procedure of the exposure task. With this, we at least provide the information required to opt out of the study, e.g., if they are not comfortable with using LLMs.
Finally, we inform participants about the data we collect and ask for their consent.

\subsubsection{Collection of Demographic Information}
We ask participants about their professional programming experience in years (counting part-time jobs by half). We also ask them to rate their general programming proficiency, Java proficiency, and their experience and frequency of working with LLMs or LLM tools on a scale of 1 (low) to 5 (high). We treat these variables as confounders. As we expected a relatively small sample size, we use a biased coin method (please refer to the replication package for details and the resulting distribution) to assign each participant, balanced with regard to the confounders, to one of the six possible pairs of modes and tasks (\(3! = 6\)), i.e., each mode is paired with exactly one task. We randomize the order of tasks during the programming phase and use the same order during the comprehension phase to keep the duration between exposure and comprehension as similar as possible for each task.
This yields \(3! \times 3! = 36\) possible experiment flows.

\subsubsection{Exposure Phase}

For each task, participants are presented with the task description before starting to solve the task. 
Only after participants declare that they feel ready to start programming, we inform participants whether or not they are supposed to use the provided LLM tool, or whether they are supposed to solve the task at all. In this way, we prevent participants from altering their reading and comprehension efforts by anticipating the way of (not) solving a task at least for the first two exposure tasks. 

We take screen recordings and measure the time that participants work on a task. We allow participants to use any online resource (search engines, forums, etc.) or standard programs (e.g., calculator) they like, limiting only the usage of LLMs. Once they are confident that they have solved a task, they can submit their solution for review. We then use the prepared test suite to check the produced code, securing some level of correctness. If the tests fail, participants are tasked with revising their solution. However, they are provided with a hint about a failed test case in the form of an expected input-output pair compared to their actual result. This procedure is meant to decrease the frustration a participant may experience due to repeated incorrect solutions. We only measure the time in which participants work on their solutions, i.e., we stop the measurement while testing their solutions or providing feedback.

\subsubsection{Interview -- Part I}
After participants completed all exposure tasks, we conduct the first part of the interview which is audio-recorded.
For each of the exposure tasks, we ask participants to rate their understanding of the task description and their understanding of the solution they produced on a scale of 1 (low) to 5 (high) and give reasons for their rating.
We also ask them how they made sure that their code is correct before submitting it in the exposure task and whether they want to tell us anything else about the experiment or their work on the tasks so far.

\subsubsection{Comprehension Phase}\label{subsubsec:comp_task}
Each task is displayed in the experiment application with the task description, an exemplary method invocation, and either the code that the participant produced (for the modes \textsc{self} and \textsc{llm}) or the pre-written solution (for the mode \textsc{unseen}). 
Each piece of code has four mutations, except for the \textsc{unseen} code, which has five mutations. We did this to prevent participants from ``exploiting'' an expected number of mutations.

For the comprehension task, participants are not allowed to use any additional resources and are required to complete the task using our application. We measure the time it takes for participants to submit each task. After each task, participants have the opportunity to take a short break to switch mental contexts and prepare themselves for the next task. 

\subsubsection{Interview -- Part II}

In the second interview, we review all mutations marked by participants and ask them whether they marked a particular mutation because they recalled the code from memory or whether they had to think about it. At this point, we also start the debriefing process as we tell them whether their marked mutations were correct or not, and whether they have missed any. 

Afterwards, we also reveal the actual goal of the study and ask them whether they feel that their comprehension of code is influenced by using GenAI. We also ask if they feel that those differences are important to them, and if so, how they counteract them. 

We conclude the interview with a final debriefing, thanking participants for their time and reassuring them of their performance. We ask them not to share the real goal of the study with other people to prevent biasing future participants. We also offer insights into preliminary results and to notify them on publication.

\subsection{Analysis Procedure}
In the following, we describe our analysis procedure for each research question.
\subsubsection{\ref{rq1}}
We first use descriptive statistics to visualize the data set. Afterwards, we employ a linear mixed-effects model (LMEM) and follow the recommendations of \textcite{vegas2016}. Thus, we account for repeated measurements nested within the participants in the form of a random intercept, as well as ordering effects and effects associated with the concrete tasks. We also consider the participant-specific variables for the model such as programming experience, the different proficiency variables and whether the experiment was conducted locally or remotely. We use \textit{seconds\_per\_mutation} as response variable, i.e., how much time a participant spent on average to find one introduced mutation. This variable, therefore, combines the measures of time taken for each comprehension task and the number of correctly identified mutations. In case a participant finds none of the introduced mutations, we assume the total time spent on the respective comprehension task as \textit{seconds\_per\_mutation}.
In our LMEM, we omit the interactions between mode, order, and task as we do not expect an asymmetric carryover.
We use repeated contrasts for the categorical variables \textit{mode} and \textit{task} as described by \textcite{SCHAD2020104038}, e.g., we construct the contrast for \textit{mode} to compare the \textsc{unseen} level to the \textsc{llm} level and the \textsc{llm} level to the \textsc{self} level. To compare all three modes, we conduct pairwise \(t\)-tests and adjust the \(p\) values with the Holm-Bonferroni method.  

For constructing and fitting the model, we use R with the lme4 library\footnote{\url{https://github.com/lme4/lme4}}. We conduct significance testing by using the lmerTest library~\cite{kuznetsova} and obtain degrees-of-freedom for t-tests using the Kenward-Roger method~\cite{kenward} that is designed for small sample sizes. The interpretation of our results is based on a significance level of $\alpha = 0.05$.
To validate our model, we ensure the normality of residuals and the absence of multicollinearity. In case of non-normal residuals, we log-transform \textit{seconds\_per\_mutation}. 
We finally report marginal means for the three modes, which show \textit{seconds\_per\_mutation} values that are averaged over fixed and random effects.

\subsubsection{\ref{rq2}} For the second research question, we focus on the self-reported comprehension rating of participants and its relation to \textit{seconds\_per\_mutation}. We thereby aim to see whether our methodology of measuring code comprehension aligns with the participants' ratings. For this, we follow the same approach as described for RQ1. We use a similarly constructed LMEM, adding only the variables of \textit{solution comprehension} and \textit{description comprehension} rating as fixed effects. As participants did not rate solution comprehension for the \textsc{unseen} mode, we only consider the modes \textsc{llm} and \textsc{self}.

To find out how participants reflect on possible differences in code comprehension when (not) working with LLM tools, we analyze the interview recordings. We transcribe the interview recordings using noScribe\footnote{\url{https://noscribe.de/en/}} and follow the six-step procedure of thematic analysis as described by Braun and Clarke~\cite{Braun01012006}, using an inductive, codebook-based approach. That is, we first familiarize ourselves with the interviews and follow up with an initial manual coding. In the case that codes emerge in a later stage of the initial coding, we revisit earlier coded interviews. We then search for themes related to our research question by categorizing the codes. Finally, we review and define the themes before reporting. The thematic analysis is done by one author and discussed with all authors afterwards.



\subsection{Deviations}\label{subsec:deviations}
Unfortunately, deviations from the study procedure could not be avoided. Originally, we recruited 38 participants. Three participants did not finish the experiment and were excluded from analysis.
Due to audio recording problems, we were unable to include two interviews in the respective analysis. In two other cases, the experimenter was able to take notes instead.
We were left with 35 participants for answering \ref{rq1} and 33 for answering \ref{rq2}.


\section{Results}\label{sec:results}
We present the results of our study in the following.
\subsection{RQ1: Measured Comprehension Differences}\label{subsec:comp_diff}

We started the data analysis with analyzing \(n=35\) to investigate differences in code comprehension in dependence of the respective mode. 

As a first measure, we look into the percentage of found mutations. On average, participants found 88.5\% of the introduced mutations in the \textsc{self} mode, 80\% in \textsc{llm} mode, and 66.9\% in \textsc{unseen} mode. Participants took on average 2\,min\,3\,s until submitting the comprehension task of the \textsc{self} mode, 2\,min\,33\,s for the \textsc{llm} mode, and 4\,min\,8\,s for the \textsc{unseen} mode.
\begin{figure}[t]
    \centering
    \includegraphics[width=0.85\linewidth]{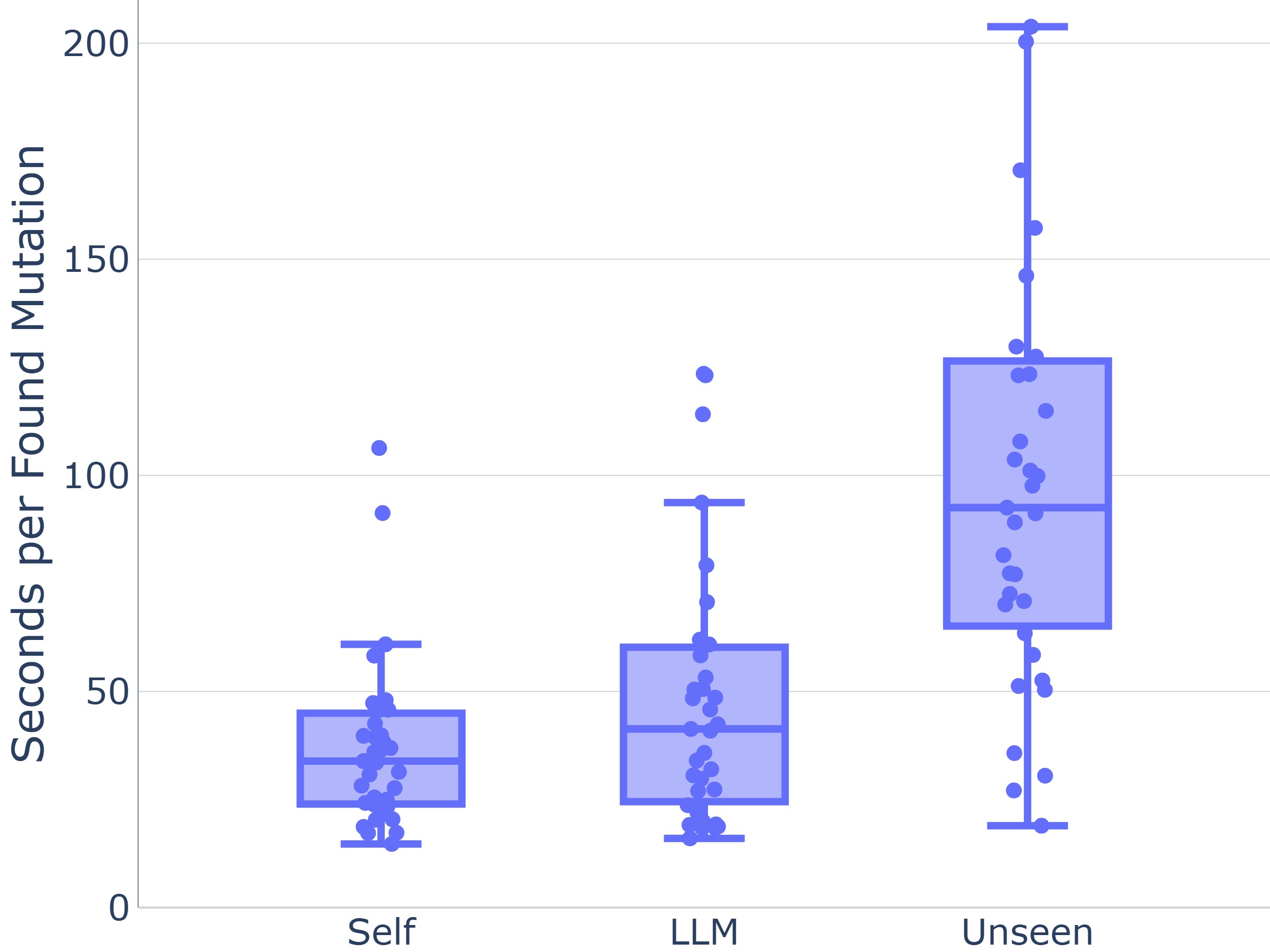}
    \caption{Seconds for finding an introduced mutation grouped by the modes. Three data points (1 \textsc{llm}, 2 \textsc{unseen}) are above 300\,s per mutation and are not visible in the figure}
    \label{fig:seconds_per_mutation}
\end{figure}
A combination of these measurements can be seen in \Cref{fig:seconds_per_mutation} which shows \textit{seconds\_per\_mutation} grouped by the different modes. On average, participants took 37.1 seconds per mutation for the \textsc{self} mode, 60.4 for the \textsc{llm} mode, and 108.5 for the \textsc{unseen} mode. 
We also asked participants for each of their marked mutations, whether the particular mutation was found due to remembering the original source code or whether it was due to something else. On average, 59.3\% of the mutations found in the \textsc{self} mode were reported as ``remembered'' compared to 34.3\% of the mutations found in the \textsc{llm} mode.
The share of incorrectly marked mutations was similarly low for the \textsc{self} (2.4\%, 3 of 127 markings) and \textsc{llm} modes (3.4\%, 4 of 116), but considerably higher for the \textsc{unseen} mode (12.7\%, 17 of 134).

\begin{figure}[t]
    \centering
    \includegraphics[width=0.85\linewidth]{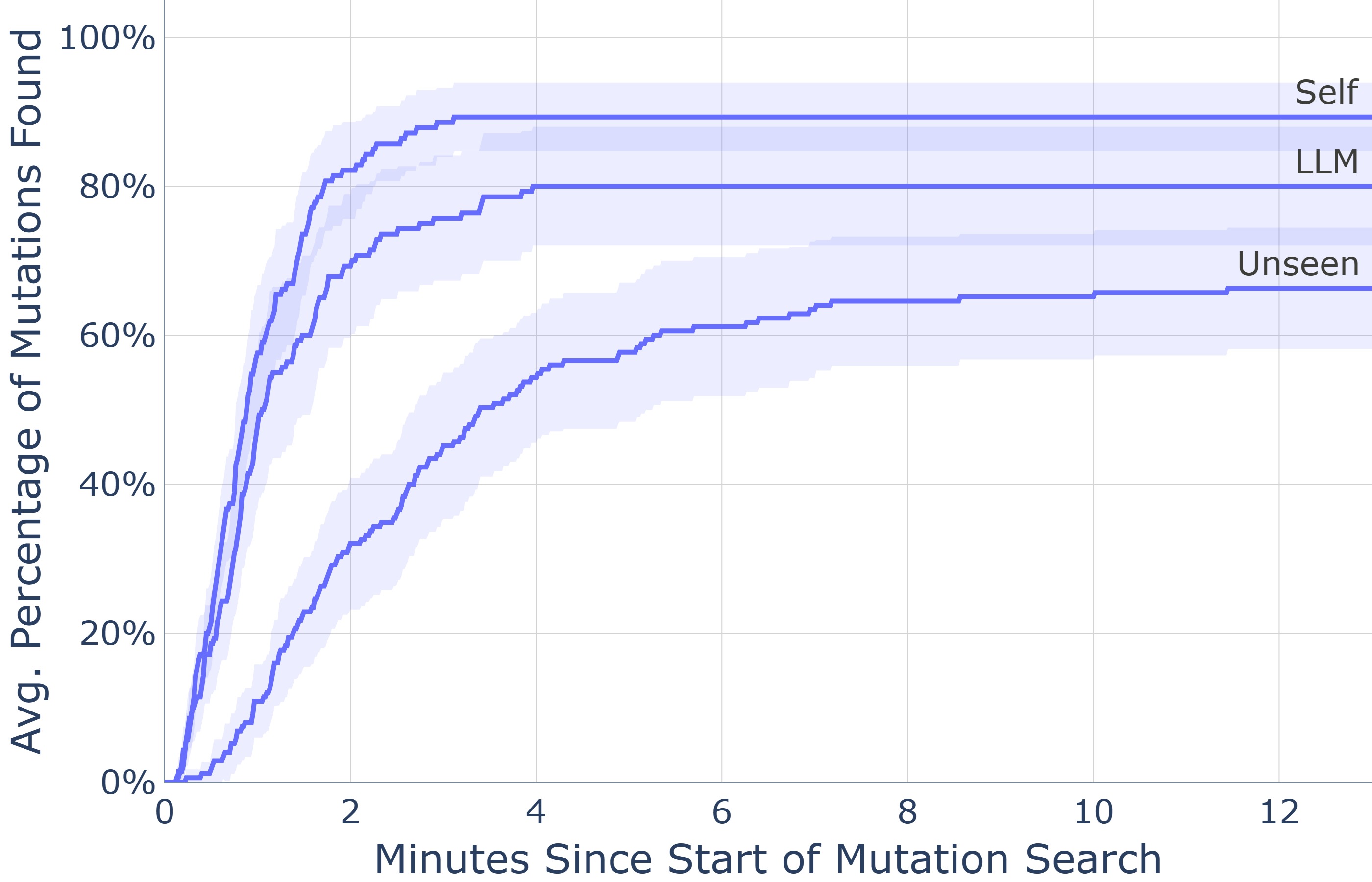}
    \caption{Average percentage of mutations found by the participants over time split by modes. Shaded areas represent 95\% confidence intervals. Note that \textsc{unseen} cannot be compared exactly as in this mode there were five mutations to be found in comparison to the four mutations for other modes.}
    \label{fig:percentage_mutations_found}
\end{figure}
In summary, participants needed less time when looking for mutations in self-written code and found more mutations compared to code produced with an LLM. Additionally, participants performed better according to the same metrics when looking for mutations in generated code compared to unknown code. This is also visible in \Cref{fig:percentage_mutations_found}.
In general, recollection of the code played an important role in finding mutations. 

For the LMEM we had to log-transform \textit{seconds\_per\_mutation} in order to achieve normality of residuals. The final model did not suffer from multicollinearity.
Looking at \Cref{tab:lmm_seconds_per_mutation}, we can see that the difference between modes is also statistically significant. Comparing the \textsc{llm} mode to the \textsc{unseen} mode, we get a 52.3\% reduction in \textit{seconds\_per\_mutation} with \(p<0.001\). Similarly, the \textsc{self} mode receives a 21.7\% reduction in \textit{seconds\_per\_mutation} with \(p=0.043\) compared to the \textsc{llm} mode. The estimated marginal means are 87.2\,s [72.2, 105.2] for \textsc{unseen}, 41.6\,s [34.4, 50.1] for \textsc{llm}, and 32.5\,s [27.0, 39.3] for \textsc{self}. This means that participants took 268\% longer to find a correct mutation in the \textsc{self} mode compared to the \textsc{unseen} mode. All Holm-adjusted pairwise comparisons were significant. 

Interestingly, we also see a significant influence of the participants' professional programming experience and their overall coding proficiency. In our sample, more experience slightly decreased the performance in the comprehension phase, while a higher coding proficiency is also associated with a higher performance. Java proficiency, LLM proficiency, order, remote participant, and differences in tasks were not significant. 


\begin{table}[t]\centering
\scriptsize
\caption{Results of the LMEM for Seconds per Mutation based on 105 observations. The first rows show the application of repeated contrasts on the variables mode and task. Columns:\(\beta\): estimated coefficients, SE: standard error, \(t\): value of the \(t\)-statistic, \(p\): \(p\)-value, 95\% CI: Confidence intervals at \(\alpha=0.05\), \% Change: Percentage increase/decrease when moving one point on the respective scale (e.g., one more year of experience increased \textit{seconds\_per\_mutation} by approx. 5\% in our sample.)}\label{tab:lmm_seconds_per_mutation}
\setlength{\tabcolsep}{5pt}
\begin{tabular}{lrrrrrrr}\toprule
Term &\(\beta\) &SE &\(t\) &\(p\) &95\% CI & \% Change \\\midrule
\textsc{llm}-\textsc{unseen} &-0.741 &0.118 &-6.27 &<.001 &[-0.98, -0.51] &-52.3 \\
\textsc{self}-\textsc{llm} &-0.245 &0.119 &-2.06 &.043 &[-0.48, -0.01] &-21.7 \\
T2-T1 &-0.157 &0.118 &-1.32 &.191 &[-0.39, 0.08] &-14.5 \\
T3-T2 &0.066 &0.119 &0.55 &.583 &[-0.17, 0.30] &6.8 \\
order &-0.090 &0.059 &-1.53 &.132 &[-0.21, 0.03] &-8.6 \\
is\_local &0.167 &0.137 &1.22 &.231 &[-0.11, 0.45] &18.2 \\
exp &0.052 &0.017 &2.99 &.006 &[0.02, 0.09] &5.3 \\
coding\_p &-0.467 &0.103 &-4.54 &<.001 &[-0.68, -0.26] &-37.3 \\
java\_p &0.132 &0.074 &1.80 &.082 &[-0.02, 0.28] &14.2 \\
llm\_p &0.062 &0.071 &0.87 &.391 &[-0.08, 0.21] &6.4 \\
\bottomrule
\end{tabular}
\end{table}

\begin{ObservationBox}
\textbf{Answer to \ref{rq1}}: Developers gain a worse understanding of code produced with LLM assistance compared to self-written code. However, they demonstrate a better understanding compared to unknown code. We find significant differences in our key metric \textit{seconds\_per\_mutation} with developers taking 87.2 seconds per mutation for unknown code, 41.6 seconds for code produced with LLM assistance, and 32.5 for self-written code.

  \label{obs1}  
\end{ObservationBox}

\subsection{RQ2: Developer Perspectives on Comprehension}\label{subsec:dev_perspective}
In this section, we investigate developers' reflection on comprehension differences, drawing on self-reported comprehension ratings during the experiment and supplemented by insights from the thematic analysis of the interviews.
\subsubsection{Self-rated Comprehension of Experiment Tasks}
\begin{table}[t]\centering
\scriptsize
\caption{Results of the LMEM for \textit{seconds\_per\_mutation} based on 66 comprehension ratings. For legend refer to \Cref{tab:lmm_seconds_per_mutation}}\label{tab:self_rated_comprehension}
\begin{tabular}{lrrrrrrr}\toprule
Term &\(\beta\) &SE &\(t\) &\(p\) &95\% CI & \% Change \\\midrule
solution\_p &-0.218 &0.104 &-2.09 &.041 &[-0.43, -0.01] &-19.6 \\
description\_p &0.022 &0.090 &0.25 &.805 &[-0.16, 0.20] &2.2 \\
\textsc{self}-\textsc{llm} &-0.162 &0.145 &-1.11 &.273 &[-0.46, 0.13] &-15.0 \\
exp &0.033 &0.019 &1.70 &.101 &[-0.01, 0.07] &3.3 \\
coding\_p &-0.416 &0.115 &-3.61 &.001 &[-0.65, -0.18] &-34.0 \\
java\_p &0.140 &0.083 &1.69 &.102 &[-0.03, 0.31] &15.0 \\
\bottomrule
\end{tabular}
\end{table}

\begin{table}[t]
  \centering
  \caption{Self-reported comprehension ratings (\(n=33\)). Each cell gives the median rating, followed by the number of participants who chose rating 1/2/3/4/5. Code comprehension was not collected for \textsc{unseen}.}
  \label{tab:comprehension}
  \begin{tabular}{lccc}
    \toprule
                & \textsc{self}       & \textsc{llm}         & \textsc{unseen}      \\
    \midrule
    Description & 4 \;(0/1/7/10/15)   & 4 \;(0/2/2/17/12)    & 4 \;(2/2/3/10/16)  \\
    Code        & 5 \;(0/0/0/9/24)    & 5 \;(1/0/5/10/17)    & ---                  \\
    \bottomrule
  \end{tabular}
\end{table}
After completion of the exposure phase, we asked participants to rate their comprehension of each individual task description and their comprehension of the code they produced in the \textsc{llm} and \textsc{self} modes. From this, we obtained the ratings of \(n=33\) participants, which are shown in \Cref{tab:comprehension}. 
In 18 cases, participants reported the same comprehension rating for \textsc{self} mode and \textsc{llm} mode, in 12 cases participants reported a higher rating for \textsc{self} than \textsc{llm}, and in 3 cases a higher rating was reported for \textsc{llm} compared to \textsc{self}.
Overall, the participants' comprehension rating of generated code is slightly lower than their comprehension rating of self-written code, with around half of the participants reporting no differences.

\Cref{tab:self_rated_comprehension} shows the influence of the two comprehension ratings on the \textit{seconds\_per\_mutation} metric.
Again, we had to log-transform \textit{seconds\_per\_mutation} in order to achieve normality of residuals. The final model did not suffer from multicollinearity. Compared to the model presented in \Cref{tab:lmm_seconds_per_mutation}, we had to remove some of the fixed effects to account for the now reduced sample size, as we are not considering the mode \textsc{unseen}. We removed variables with \(p>0.1\).

According to the LMEM, the reported solution comprehension is a significant predictor (\(p\)=\(0.041\)) of \textit{seconds\_per\_mutation}. A one-step increase of this rating reduced \textit{seconds\_per\_mutation} by 19.6\%. A higher comprehension rating of the description also reduced \textit{seconds\_per\_mutation}. However, the influence was not significant (\(p\)=\(0.805\)). There are no other significant variables. Note that in this case, the different modes had no significant influence. This is likely caused by a reduction in observations, as we could not consider the \textsc{unseen} mode in this model and a partial mediation through the comprehension rating.

In summary, the self-reported comprehension rating seems to align with our comprehension proxy \textit{seconds\_per\_mutation}, i.e., a higher self-reported comprehension rating correlates with a lower time spent per mutation. However, only about half of participants report comprehension differences between programming with LLMs and programming themselves.

\subsubsection{Perception of Comprehension of LLM-generated Code}
The following represents the results of the thematic analysis of \(n=33\) interviews.
\paragraph{General Comprehension Differences}
After the experiment, we asked participants about their general perspective on the comprehension of generated code compared to self-written and unknown code. All of the participants agreed that in general, they do not have the same degree of understanding of generated code in comparison to self-written code. Additionally, nine participants acknowledged that they usually have a weak understanding of generated code and that it feels more like unknown code to them: \textit{``It did not feel that different. Both were simply code that I had to understand from scratch''}. Three participants were certain that one cannot gain the same understanding of generated code compared to self-written code even when putting in the effort to do so especially for more complex code.
In contrast, three participants saw no or only negligible differences for tasks as small as in this experiment but expected them to grow with the size of the task (\textit{``I would say I understand the code about equally well. [\ldots] But only with this limitation: it is a small task. [\ldots] If it scales up this changes quickly.''}).

\paragraph{Comprehension Differences in the Experiment}
In relation to the comprehension phase of the experiment, five participants believed that making errors in the exposure task significantly helped them to understand possible pitfalls in the code and to gain an intuition on where to look for mutations (\textit{``[I had] already made the mistakes, and then you knew what to look out for.''}). Two others reported to have felt reluctant in the comprehension phase, as they could not remember their original code due to copying generated solutions. 

\paragraph{Explanation of Comprehension Differences}
Although we did not ask for it directly, several participants provided insight into why they think they had a poorer understanding of the generated code. Eight (mostly more novice) participants attributed these differences to a clash between the expected code and the generated code, for example because it was different from what they (or another person) would have done with unknown functions, ``syntactical sugar'', or it was too complex and detailed. One participant argued that they usually have an easier time understanding human-written code. They exemplified this point by referring to the comprehension phase where they had an easier time understanding the \textsc{unseen} code compared to their generated code. However, in this case the \textsc{unseen} code was generated, too.
Another participant felt like \textit{``looking at the [generated] code from an outsider perspective''}. Two participants explicitly explained that for them the main factor in understanding code is the process of writing that code and re-iterating on it, which they usually do not experience in a similar way when working with LLMs.

\paragraph{Dealing with Comprehension Differences}
Strategies for dealing with comprehension differences varied strongly between the participants. An important factor seemed to be the task at hand. Many participants (11) generally felt more comfortable off-loading programming tasks to the LLM when a task is less critical, less complex, or easily verifiable (\textit{``If I need some function and I know it has to work in a certain way, I have to pay more attention when prompting [\ldots] and, if necessary, I would rather code it myself. But for simpler things [\ldots] I just prompt and use the solution from the LLM more quickly, because I know that not much can go wrong.''}). In these cases, participants were also more likely to accept a lower understanding of the code. 
However, seven participants said that when working on hobby projects or when learning something new, they avoid using LLMs for coding to maintain control of their projects and develop a stronger understanding of new concepts (\textit{``I notice for myself that it has a considerable negative effect. Because if I don't do things myself, I don't learn them properly.'}'). 

Six participants explained that due to comprehension differences, they invest more time in understanding solutions to alleviate these differences. Another seven participants instead opted to reduce the amount of code generated at once by adjusting their prompts to only get smaller portions of the desired code (\textit{``I have moved more towards only having small nuggets
generated.''}). 

Six participants said that they are actively using LLMs to help them understand code or parts of code. This could also be the case for generated code where they would prompt for an accompanying explanation of the code.


\paragraph{Risks of Generating Code}
Six participants reported the problem of losing themselves in feedback loops without resolution, meaning that even after several iterations with additional context (such as error messages), the LLM they used was unable to generate a correct one (\textit{``[\ldots] you get
into a spiral: you paste in generated code, it is executed, it doesn't work, you feed in the error message, you get new code, paste it in, it doesn't work.''}). In these cases, the participants felt frustrated with the amount of time invested without usable output or a better understanding of the problem at hand.
Ten participants also stated that they had checked the generated code in the experiment only superficially (\textit{``For the second one, I just said: tests pass, zap, done. That's a problem for my future self.''})

\begin{ObservationBox}
\textbf{Answer to \ref{rq2}:} The self-reported comprehension rating of developers was aligned with our measurement of code comprehension according to the LMEM. Developers generally reported a lower comprehension of generated code in comparison to self-written code. How developers address these issues varies based on the individual developer and the concrete task at hand. Some developers tend to accept comprehension differences in uncritical, low-complexity, or easily verifiable code.
  \label{obs2}  
\end{ObservationBox}


\subsection{Additional Observations}
This section bundles exploratory analyses which we did not plan for beforehand.
\subsubsection{Verification Activities}
A lower understanding of solutions might result in higher verification efforts by participants, because they saved time in the production of code that they can now invest in more thorough testing. To investigate this, we systematically reviewed screen recordings of \(n=35\) participants to identify verification activities, which are presented in \Cref{tab:verification}.
We find no obvious differences in executed verification activities between the two modes.
 

\begin{table}[t]
    \centering
    \caption{Verification activities extracted from the screen recordings of 35 participants}
    \begin{tabular}{lrr}\toprule
        Activity & \textsc{SELF} & \textsc{LLM} \\\midrule
         Execute main method & 26 & 25 \\
         Change main method & 13& 8\\
         Execute test suite & 25& 27\\
         Extend test suite & 5& 6\\
         Use debugger & 4&0 \\
         Re-read description & 7&11 \\
         Prompting & 0& 2\\
         \bottomrule
    \end{tabular}
    \label{tab:verification}
\end{table}
\subsubsection{Programming Time}
We observed a stark contrast in the required programming time between programming with (Avg. 14\,min\,31\,s) and without LLM (Avg. 24\,min\,41\,s). We build another LMEM with programming time as the dependent variable to investigate the impact of the concrete mode in detail. Based on the model, we find that programming without LLM increased the programming time by 91.3\% compared to programming with LLM (\(\beta=0.65\), \(SE=0.118\), \(t=5.48\), \(p<0.001\), CI:\([0.41,0.89]\)).

\subsubsection{Impact of LLM Usage on Comprehension in \textsc{llm} mode}
The results of \textcite{Alakmeh2026} suggest that the degree of comprehension that a developer gains during the usage of LLMs depends on how the LLMs are used. To investigate this notion in our data, we introduce two new variables \texttt{gen\_code\_used} and \texttt{desc\_containment} as proxies for reliance on the LLM and engagement with the task, respectively. With \texttt{gen\_code\_used} we measure the percentage of code lines in the final solution that were also part of the LLM responses. With \texttt{desc\_containment} we measure the percentage of words that were copied from the task description to the prompts. We build a linear regression model with \texttt{log(seconds\_per\_mutation)} as the dependent variable and find that LLM proficiency (\(\beta=0.30\),\(p=0.017\)), \texttt{gen\_code\_used} (\(\beta=0.81\),\(p<0.05\)), \texttt{desc\_containment} (\(\beta=0.73\),\(p=0.002\)), and the interaction term between the two new variables (\(\beta=2.49\),\(p=0.002\)) have a significant influence.

\section{Discussion}\label{sec:discussion}
Our study points to two key findings:
\paragraph{Cognitive debt is real, but less pronounced for small tasks} Participants found mutations faster and more reliably in self-written than in LLM-produced code. This provides evidence for the claims of \textcite{storey2026} that writing code builds a mental model as a by-product, while accepting generated code may not. As both original solutions passed our prepared test suites, this also strengthens the claim that cognitive debt can accumulate behind working code and passing tests~\cite{ahmad2026, tavantzis2026cognitivedebtgenaiera}. Participants also recalled less of their generated code which was also observed by \textcite{Kosmyna2025} for LLM-written essays. At the same time, generated code was understood significantly better than unseen code, indicating that prompting, judging, and integrating the output builds at least a partial understanding. The programming time was nearly halved and the debt in our setting was small and promptly repaid, as described by \textcite{Cunningham1992}. Nevertheless, participants missed more mutations in the \textsc{llm} mode. Cognitive debt, therefore, may be paid not only in time but also in the quality of work.
\textbf{Implication for Researchers:} Further research is warranted on how cognitive debt accrues interest, whether it accumulates over time and tasks~\cite{chou2026vibeguidegraphbaseduserinterface}, and how it behaves for programming activities in larger systems.

\paragraph{Developers are aware of comprehension differences} We find that our participants' self-reflection of code comprehension was generally aligned with our measurement of code comprehension. 

The way our participants deal with comprehension differences varied between participants and is task-dependent. Accepting the differences for uncritical, low-complexity, or easily verifiable code seemed to be widely accepted. However, for some participants, this was the main reason not to use LLMs for programming or to drastically reduce the amount of code generated at once.
This can also be seen in our quantitative results, where less reliance on generated code and formulating own prompts predicted faster mutation finding.
\textbf{Implication for Researchers:} 
Decisions about delegation seem to be made according to a concrete task or goal, where cognitive debt can be seen as a risk or strategy~\cite{storey2026}. Future work should therefore examine how well developers' judgments hold up.
\textbf{Implication for Practitioners:} It might not be obvious whether a (future) task requires a more in depth understanding of a system or piece of code. Whether work is offloaded to LLMs should therefore be evaluated carefully and consciously. 


\section{Threats to Validity}\label{sec:threats} 
Our study is subject to some aspects that threaten the validity and generalizability of the results. We distinguish between construct, internal, external, and conclusion validity~\cite{wohlin2024experimentation}. 

\subsection{Construct Validity}
For our study, we used different problem descriptions to avoid the mono-operation bias. However, the different tasks may have affected the construct validity, as the goal of the exposure task does not align with the goal of the comprehension task. Based on the briefing, we expected participants to focus on code correctness and completion time, i.e. producing correct code in the shortest amount of time. However, producing source code and finding errors in source code does not necessarily require the same understanding of the program. Therefore, skills required to comprehend generated code or code to be written may differ from skills required to comprehend existing code. Nonetheless, we argue that the understanding required for each task is connected and that advancing the understanding as required for one task also benefits the understanding of the other task in our experiment.

In addition, we cannot be sure that finding mutations actually measures code comprehension. We have seen that participants remembered more of the self-written code than of the generated code, which points to the assumption that we rather measured the ability to recall the code. However, in \Cref{subsec:dev_perspective}, we see that self-reported code comprehension has a significant influence on \textit{seconds\_per\_mutation}, hinting at a connection between both measurements. As both memory and comprehension (wherever the line between the two is drawn) feed into cognitive debt, we argue that this does not invalidate our results.

Further, our results may be subject to the Rosenthal effect: especially following the comprehension task, participants may have thought that the experimenter is trying to find differences in comprehension of self-written and LLM-generated code. This may have influenced their answers to draw a picture that they think suits better to what the experimenter wants to find. 

\subsection{Internal Validity}
The internal validity may be threatened due to unfairness with regard to some variables we used. For example, when comparing the mutations found per second, the calculation may be somewhat unfair as in the \textsc{unseen} mode, there were five mutations to be found, while the other modes had only four mutations. 


\subsection{External Validity}
The participants in our experiment come from industry and academia. They have at least basic programming skills with Java and are undergraduate, graduate or PhD students, or professional software developers. We recruited participants using convenience sampling, which strongly limits generalizability. 
However, in terms of knowledge and experience levels, we have an at least somewhat heterogeneous group of participants.

Due to the artificial environment and the experimental setting, the external validity of our results may be threatened. Future studies should hence triangulate our results with insights from real software projects, for example, using action research or observational studies. 

We only considered the use of a chat as LLM tool, which is only one scenario. In fact, developers may also use agentic or enhanced auto-completion which is focused on interaction. These usage scenarios are not covered by our experiment. 

In addition, the concrete LLM used for the programming phase influences the probability of generating correct solutions, lowering the amount of manual intervention and re-prompting by developers and in turn possibly increasing cognitive debt. However, another scenario is also possible: the output of a more powerful LLM might align better with the expectations of developers, and thereby reduce cognitive debt.

\subsection{Conclusion Validity}
The results of our experiment are based on a sample size of at most $n = 35$ participants. Hence, they should be carefully interpreted. To draw reliable conclusions, they are based on quantitative data analyzes that comprise different (regression) models and statistical tests. We relied on established best practices as described by Vegas et al.~\cite{vegas2016} to analyze our data taking into account the special setting. 

\section{Conclusion}\label{sec:conclusion}
With this study, we present insights into the notion of cognitive debt and show that there exists a trade-off between speedup through the use of LLM tools with a subsequent slowdown in comprehension activities such as bug finding. We show that developers are generally aware of their weaker understanding of generated code and present strategies to deal with these comprehension differences. 
Our results show promise for further research on cognitive debt and the perspectives of developers of it.

\section{Data Availability}
Please refer to \url{https://tinyurl.com/comprehension-llm-code} for the replication package containing the material needed to conduct the study, as well as results and analysis scripts.

\section*{Acknowledgments}
We thank our participants for taking part in the study and providing us with valuable insights.

We used LLM tools to support programming activities related to the study, including the creation of the application used to guide through the experiment, as well as the code used to analyze and create figures. We used GitHub Copilot in PyCharm for enhanced auto-completion and utilized Gemini, ChatGPT, and Claude Opus 5.5 for more complex code generation and debugging.
\balance
\printbibliography

\end{document}